%% file: tplp09.tex
\documentclass{tlp}
\usepackage{epsf,epsfig}
\usepackage[usenames]{color}  % using for comments during paper writing.
\newcommand\bcmdtab{\noindent\bgroup\tabcolsep=0pt%
  \begin{tabular}{@{}p{10pc}@{}p{20pc}@{}}}
\newcommand\ecmdtab{\end{tabular}\egroup}

\newcommand\tls[1]{{\sc \textcolor{red}{#1}}}
\newcommand\dsw[1]{{\sc \textcolor{CornflowerBlue}{#1}}}
\newcommand\comment[1]{}

\newcommand{\cF}{{\cal F}}

\newcommand{\cO}{{\cal O}}

\newcommand{\cS}{{\cal S}}

\newcommand{\mif}{\mbox{ :- }}

\newcommand{\version}{Version 3.3}

  \title[The XSB System]
        {XSB: Extending Prolog with Tabled Logic Programming}

  \author[Terrance Swift and David S. Warren]
         {TERRANCE SWIFT\\
           CENTRIA, Faculdade de Ci\^{e}ncias e Tecnologia \\ 
           Univ. Nova de Lisboa, 2825-516 Caparica, Portugal \\
         \email{tswift@cs.sunysb.edu}
         \and DAVID S. WARREN \\
         Computer Science Department, 
         SUNY Stony Brook \\
         Stony Brook, New York, USA \\
         \email{warren@cs.sunysb.edu}
}

\submitted {4th October 2009}
\revised {28th February 2010}
\accepted{22nd November 2010}

\newtheorem{example}{Example}[section]

\begin{document}

\label{firstpage}

\maketitle

\input{abstract}

% \begin{keywords}
% Prolog, logic programming system
%\end{keywords}


%\tableofcontents

\input{intro}

\input{tbe}

\input{dynamic}

\input{mt}

\input{applications}

\input{discussion}

\bibliographystyle{acmtrans}
\bibliography{longstring,all}

\label{lastpage}
\end{document}

%% file: abstract.tex
\begin{abstract}
The paradigm of Tabled Logic Programming (TLP) is now supported by a
number of Prolog systems, including XSB, YAP Prolog, B-Prolog,
Mercury, ALS, and Ciao. The reasons for this are partly theoretical:
tabling ensures termination and optimal known complexity for queries
to a large class of programs.  However the overriding reasons are
practical.  TLP allows sophisticated programs to be written concisely
and efficiently, especially when mechanisms such as tabled negation
and call and answer subsumption are supported.  As a result TLP has
now been used in a variety of applications from program analysis to
querying over the semantic web.  This paper provides a survey of TLP
and its applications as implemented in XSB Prolog, along with
discussion of how XSB supports tabling with dynamically changing code,
and in a multi-threaded environment~\footnote{To appear in Theory and
  Practice of Logic Programming (TPLP)}.
\end{abstract}

%% file: intro.tex
\section{Introduction}
Since its inception, a primary goal of XSB has been to expand the
areas in which Prolog is used, by making Prolog more powerful, more efficient, and more
declarative.  In 1993 when XSB was first released, it supported this
goal by including both tabled resolution for definite programs, 
which provided it with deductive database-style features of
such systems as Coral~\cite{RSS92} and LDL~\cite{CGK90}.  At the time,
while XSB was faster than those systems, it was basically suitable
only for research by its developers.  Since then, XSB has become a
widely used multi-threaded Prolog that is compliant with most
standards.  During this development, XSB's research focus has
continued to be centered on tabling.

At one level, the idea behind tabling is simple; subgoals encountered
in a query evaluation are maintained in a table, along with answers to
these subgoals.  If a subgoal is re-encountered, the evaluation reuses
information from the table rather than re-performing resolution
against program clauses.  For instance using tabling, a Prolog
predicate for transitive closure over a graph:
\begin{verbatim}
  reach(X,Y):- edge(X,Y).
  reach(X,Y):- edge(X,Z),reach(Z,Y).
\end{verbatim}
could just as easily be written as
\begin{verbatim}
  reach(X,Y):- edge(X,Y).
  reach(X,Y):- reach(X,Z),edge(Z,Y).
\end{verbatim}
(and, as discussed below, there are good reasons for performing this
rewrite).

This simple idea has profound consequences.  First, tabling ensures
termination of programs with the {\em bounded term-size property} --
those programs where the sizes of subgoals and answers produced during
an evaluation are less than some fixed number.  This makes it much
easier to reason about termination than in basic Prolog.  Second,
tabling can be extended to evaluate programs with negation according
to the Well-Founded Semantics (WFS)~\cite{VRS91}.  Third, for queries
to wide classes of programs, such as datalog programs with negation,
tabling (perhaps combined with compiler transformations) can achieve
the optimal complexity for query evaluation.  And finally, tabling
integrates closely with Prolog, so that Prolog's familiar programming
environment can be used, and no other language is required to build
complete systems.

These properties have led to the emerging paradigm of Tabled Logic
Programming (TLP).  The properties of termination and optimal
complexity have made TLP useful to explore state spaces in
applications from program analysis to process and temporal logics.
The termination properties together with WFS have supported a variety
of extensions for non-monotonic constructs such as annotations,
preferences, explicit negation, and abduction; and have led to the
integration of Prolog with Answer Set Programming (ASP) through XSB's
XASP package.  Together these properties have fostered combinations of
TLP with more declarative and less procedural knowledge representation
approaches.  
%% dsw: deleted these forward refs. Good but necessary?
%%Examples based on XSB include the object-oriented
%%deductive database Flora-2, the semantic web reasoner Silk, along with
%%several systems that combine TLP with ontologies.

This paper discusses how XSB supports TLP, and how TLP supports
applications when combined with other features of Prolog such as
dynamic code and constraints.  Section~\ref{sec:tbe} describes the
tabling features of XSB (including reclaiming table space) through a series of examples including tabling
for definite programs, tabled negation, call and answer subsumption,
and tabled constraints.
Section~\ref{sec:dyn} then describes XSB's approach to dynamic code,
including its integration with TLP via incremental tabling.
Section~\ref{sec:mt} discusses XSB's multi-threading.
% particularly its support for tables in a
%multi-threaded environment.  
Finally Section~\ref{sec:applic} discusses two applications 
that we consider particularly innovative and significant.

However before proceeding, we briefly consider XSB purely as a Prolog
system.  Each version of XSB runs on Linux, Mac OS and Windows
(compiled with either MSVC or Cygwin); for Linux and Mac OS 64-bit
compilation is supported in addition to 32-bit.  With a few
exceptions, XSB supports the core Prolog standard (ISO-IEC 13211-1),
the core revision working draft (ISO/IEC DTR 13211-1:2006) and, as
discussed in Section~\ref{sec:mt}, the multi-threading working draft
(ISO/IEC DTR 13211-5:2007).
%However, XSB's module system does not support the ISO
%standard (nor do many other Prolog systems).
%(???).  The non-compliance results from a
%design decision in XSB's module system that allows it to be very fast
%for calling meta-predicates; 
%Nonetheless, most uses of modules work identically in XSB as in other
%Prologs and 
\comment{  Furthermore, \version{} supports the uses of modules required by the
Prolog-commons, (a working group formed to enable code to be shared
more easily among Prolog systems).  
}
XSB also supports constraint logic
programming through attributed variables, the interface to which is
nearly identical to that of SWI and YAP Prolog.  As a result, constraint
libraries are ported to XSB in a routine manner, and XSB supports
Constraint Handling Rules, CLP(R) and CLP(FD).

%% file: tbe.tex
\section{Tabling By Example} \label{sec:tbe}

\comment{ \tls{No performance in here right now.  Do we put numbers
  in this section, have a separate performance section, or not even
  discuss performance at the level of tables and graphs?}  
\dsw{I'm
  not sure what to do about performance.  Raw, absolute performance
  numbers are not very helpful.  They depend on what machine is being
  used, and the fastest machine can change monthly.  So only
  comparative performance numbers are helpful.  So do we compare to
  say nontabled versions?  But those are of different complexities, so
  choose a large N and you can get any difference you want.  So, to
  repeat, I'm not sure what we might do about performance...}
}
We use a series of examples to introduce various aspects of programming with
tabling.  XSB's tabling is based on SLG
resolution~\cite{CheW96}, with extensions for call and
answer subsumption as described below.  This presentation uses a forest-of-trees 
model~\cite{Swif99b} focusing on programming aspects and
system features needed for tabling support.  Accordingly, The
presentation is informal; the full formalism of tabling
and its algorithms can be found in the references.

\subsection{Definite Programs}
\input{sld-tplp}

%\input{sld-full}

\begin{figure}[htbp] 
\centering
\epsfig{file=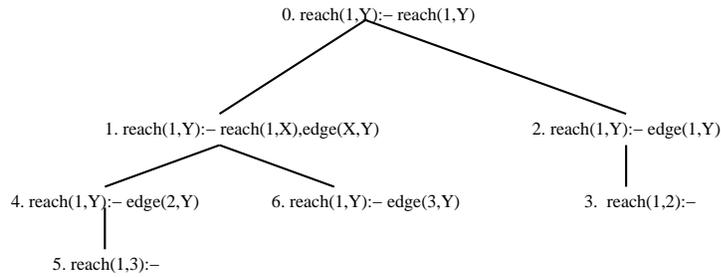,width=.8\textwidth}
\caption{Tabling tree for the query {\em reach(1,Y)} to $P_{Lrec}$}
\label{fig:deftable}
\end{figure}

%\begin{example} \label{ex:reach}
{\em Example 2.1} \label{ex:reach}
Figure~\ref{fig:deftable} shows how the XSB declaration {\tt :- table
  reach/1} affects the above program and query.
%of Figure~\ref{fig:sldtree}.
A tabled evaluation is represented as an SLG forest
% consisting of SLG trees 
in which each tabled subgoal $S$ is represented by a unique tree with
root $S \mif{} S$, which represents resolutions of program clauses and
answers to prove $S$.  (The head term conveniently collects the
bindings of subcomputations.)  The reach predicate in $P_{Lrec}$ is
left-recursive and gives rise to a single tabled subgoal, {\tt
  reach(1,Y)}, and correspondingly to a forest with a single tree.  In
Figure~\ref{fig:deftable} each non-root node of the form $K. N$ where $N = (S
\mif{} Goals)\theta$ is a clause in which the bindings to a subgoal
$S$ are maintained in $S\theta$, the goals remaining to prove $S$ are
in $Goals\theta$, and the order of creation of $N$ within the tabled
evaluation is represented by a number, $K$.  The tabled evaluation of
Figure~\ref{fig:deftable} at first resembles that of SLD: a program
clause resolves against the root node to create node 1.  However,
rather than (fruitlessly) re-applying the program clause, the
computation {\em suspends} node 1, since its selected literal has been
seen before, and uses another program clause to create node 2.  The
selected literal for node 2, {\tt edge(1,Y)}, has no table declaration
so that it is resolved as with SLD and does not create a new tree.
Program clause resolution thus creates the first answer in node 3: a
node with an empty body represents an answer.  The computation then
resumes node 1 resolving the {\em answer} against the selected literal
{\tt reach(1,X)}, and continues to derive the second answer for the
query and then to return the answer to node 1.  At that point node 6
is created and fails; no more resolutions are applicable for {\tt
  reach(1,Y)} and it is determined to be {\em completely evaluated}.
%\end{example}

%While simple, Example~\ref{ex:reach} illustrates several points.
While simple, Example 2.1 illustrates several points.
First, the evaluation keeps track of each tabled subgoal $S$ that it
encounters.  
%Later if $S$ is selected again, answers will be used to
%resolve against $S$ rather than program clauses; 
Later if $S$ is selected again, resolution will use answers 
rather than program clauses; 
if no answers are
available, the computation will {\em suspend} at that point 
%until more answers are available, 
and the evaluation will backtrack to try to derive answers using some
other computation path.  Once more answers have been derived, the
evaluation {\em resumes} the suspended computation.  Similarly, once
the computation has backtracked through all answers available for $S$
in the current state, the computation path will suspend, and resume
after further answers are found.  Thus a tabled evaluation 
is a fixed point computation for a set of interdependent
subgoals.  The second point is that by keeping a table of subgoals and
their answers, tabling can factor out redundant subcomputations --
such as the repeated SLD resolution of the selected subgoal {\tt
  reach(1,Y)}.
%in this case the repeated resolution in
%Figure~\ref{fig:sldtree} of the selected subgoal {\tt reach(1,Y)}
%using the first clause of {\tt reach/1}.  
And third, the evaluation mixes goals to tabled and non-tabled
predicates; by default predicates use SLD resolution, and only use SLG
if a {\tt table/1} declaration has been made.
%: a point discussed further in Section~\ref{}.

\begin{figure}[htb]
\centering
\mbox{
{\epsfig{file=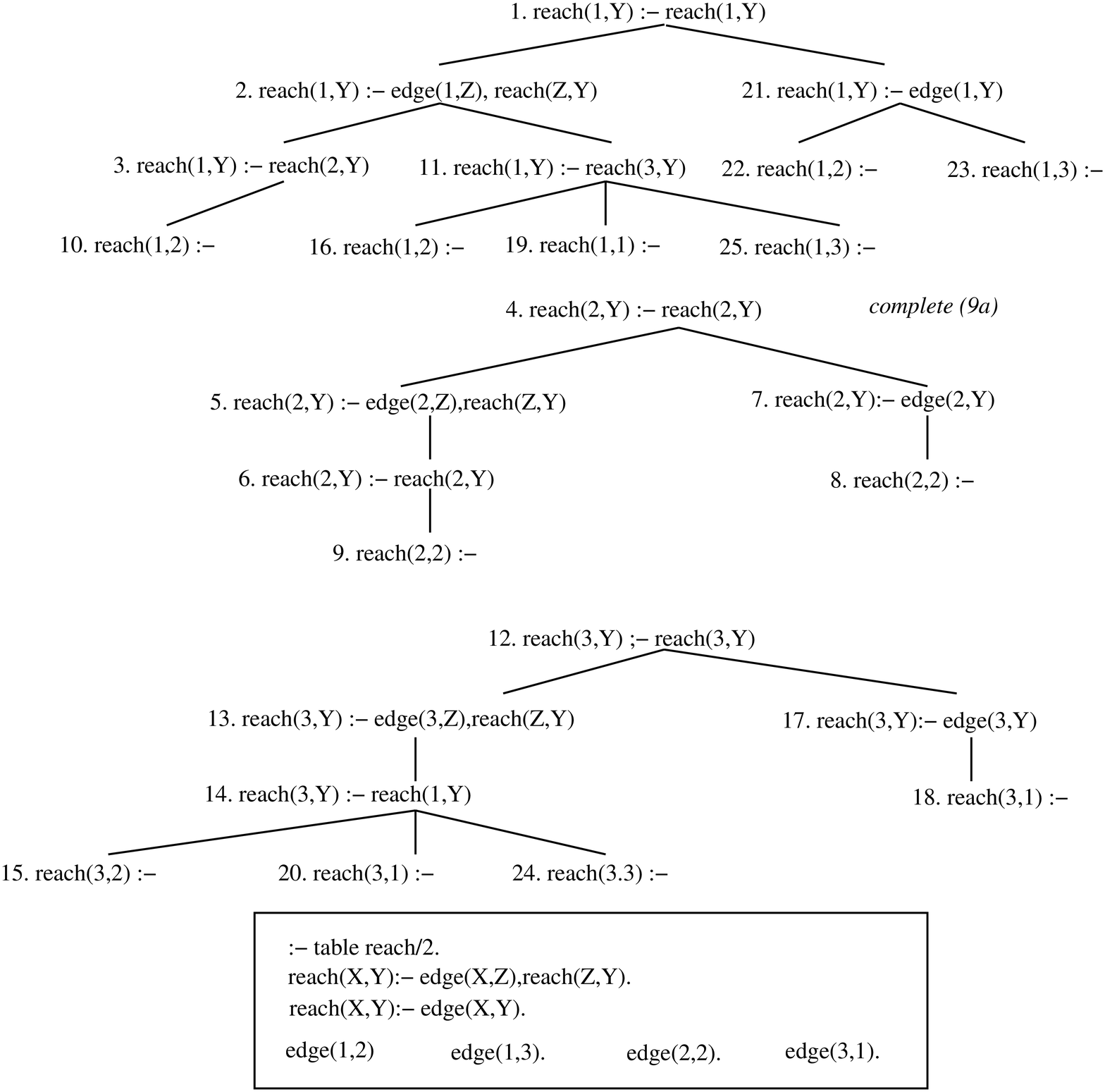,width=.99\textwidth}}}
\caption{A program $P_{Rrec}$ and SLG forest for evaluation of {\tt ?- reach(1,Y)}} \label{fig:local}
\end{figure}

At the same time, because Example~\ref{ex:reach} has only a single
tabled subgoal, it does not illustrate other important features of
tabling.  Consider for instance, the right recursive form of {\tt
  reach/1} shown in the program $P_{Rrec}$ in
Figure~\ref{fig:local}, which also shows a tabled evaluation of the
query {\tt ?- reach(1,Y)}.  There are three separate trees in
Figure~\ref{fig:local}.  At an implementation level, a tabled subgoal
together with its answers is maintained by a unique {\em table}, so 
XSB maintains three separate tables for this evaluation.
Note that the tree for {\tt reach(1,Y)} depends on {\tt reach(2,Y)} in
node 3, and on {\tt reach(3,Y)} in node 11.  Also, note that in
Figure~\ref{fig:local} the label {\em complete} is associated with the
tree for {\tt reach(2,Y)}.  If a subgoal is completed, many of its
computational resources can be reclaimed, as will be described 
in the next section.
The notion of subgoal dependency can be made precise.
%for definite programs. 
In a given forest, a non-completed subgoal $S_1$ directly depends on a
non-completed subgoal $S_2$ if $S_2$ is the selected atom of a node in
the tree for $S_1$; the definition of dependency is just the
transitive closure of direct dependency.  The direct dependency
relation for an SLG forest $\cF$ gives rise to a {\em Subgoal
  Dependency Graph (SDG($\cF$))}. Since the SDG is a directed graph, a
(maximal) set of mutually dependent goals is a strongly connected
component, or (maximal) SCC, and an independent SCC $\cS$ is a maximal
SCC such that no subgoal in $\cS$ depends on any subgoal outside of
$\cS$ (cf. \cite{MarS08}).  Note that since the SDG depends on a
forest, it changes as the forest changes, adding dependencies as new
literals are selected and deleting them as subgoals are completed.  In
Figure~\ref{fig:local}, there is one independent SCC consisting of
{\tt reach(1,Y)} and {\tt reach(3,Y)}; however in the earlier forest
that consisted of nodes 1-8, {\tt reach(2,Y)} is a trivial independent
SCC.  The importance of independent SCCs is that their subgoals can be
efficiently determined to be completely evaluated and marked as {\em
  completed} before the tabled evaluation as a whole is finished.

\vspace{-.1in}
%\subsubsection{Scheduling Strategies for Tabling} \label{sec:sched-eff}
\paragraph*{Scheduling Strategies for Tabling} \label{sec:sched-eff}
As noted, tabled evaluation has new
operations for creating a new tree, for resolving an answer against a
tabled subgoal, and for completing a mutually dependent set of
subgoals.  The order in which these operations are
applied within an evaluation is determined by a {\em scheduling strategy}.
By default XSB uses the scheduling strategy of {\em Local evaluation},
which was introduced in \cite{FrSW98} and formalized in \cite{MarS08}.
The key idea behind Local evaluation is that all operations are
performed only in a maximal independent SCC.
An alternate scheduling strategy is {\em Batched evaluation}, whose key idea is
to return an answer for a subgoal $S$ to the first node that called $S$
as soon as the answer is derived.

Local and Batched evaluation differ in that Batched evaluation eagerly
returns answers while Local evaluation may not return any answers out
of an SCC until that SCC is completely evaluated.  In general Local 
evaluation uses less stack space and is more efficient for answer subsumption
(Section \ref{sec:ans-subs}).  Batched evaluation may find first answers
faster.

% end comment to exclude batched scheduling

In addition to the decision of whether to used Batched or Local
evaluation, we mention two other principles for programming
efficiently using tabling.  First, left recursion is usually faster
for computing single-source reachability goals than other forms of recursion, such as right recursion, as left
recursion creates only a single table, and requires fewer operations.
\comment{ However, not all recursion can be transformed into (an efficient form
of) left recursion.  An example of this is the predicate {\tt sg/2},
which determines whether two nodes are in the same generation of a
DAG:
\begin{verbatim}
  sg(X,Y):- parent(X,Z),sg(Z,Z1),parent(Y,Z1).
  sg(X,X).
\end{verbatim}
For a query with mode {\em sg(bound,free)}, reordering the literals in
the first clause of {\tt sg/2} gives an immediate call with mode {\em
  sg(free,free)} which can lead to a loss of efficiency over large
DAGs.  
} % end comment
Second, tabling should be used sparingly:
for many predicates tabling will add no benefit although the table
will take up space and time to accumulate it.  In certain cases, tabling can actually increase
the complexity of a query.  For instance in XSB and all other tabled
Prologs, the query:
%\begin{verbatim}
{\tt   ?- append([a,b,c],[d,e,f],Z)}
%\end{verbatim}
to the tabled version of the normal append predicate will be quadratic
in the size of the query, as the goals {\tt
  append([a,b,c],[d,e,f],Z)}, {\tt append([b,c],[d,e,f],Z)},
etc. will be copied into the tables.

\begin{figure}[htbp] \label{fig1}
\centering
\epsfig{file=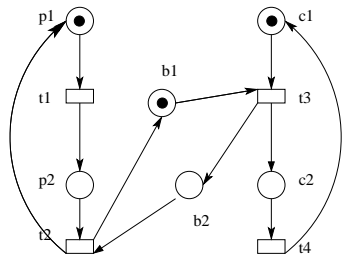,width=.35\textwidth}
\caption{A Simple Producer-Consumer Net}
\label{fig:pcnet}
\end{figure}

%\subsubsection{Example: Analyzing a Process Logic}
%
%\begin{example}[Analyzing a Process Logic]
%
{\em Example 2.2 (Analyzing a Process Logic)} The analysis of process
logics in the style of Petri Nets will illustrate various
types of tabling evaluations.
%(although other process logics could be used for this purpose).  
Reachability is a central problem for Petri
Net analysis, to which problems such as liveness, deadlock-freedom,
and the existence of home states can be reduced~\footnote{All programs
  can be obtained via {\tt
    http://xsb.cvs.sourceforge.net/xsb/mttests/benches}.}.
%While we have taken care that the programs shown are correct and
%motivated by use cases, we stress that the methods described in this
%section are intended primarily to illustrate tabling, TLP but do not
%represent fully developed analysis systems for Petri or Workflow Nets
%{\bf Using Tabling for Elementary Petri Nets}
Elementary Petri Nets (EPNs)  (cf. \cite{RosE98})
% or 1-safe Petri Nets
are particularly simple to analyze using tabling.  Consider the EPN
of Figure~\ref{fig:pcnet}, which depicts a simple producer-consumer 
system, with circles representing places and rectangles
representing transitions.
%
% Unlike a general Place-Transition Net, 
An EPN allows a place to contain at most 1 token; thus a finite EPN
has only a finite number of configurations so that determining
reachability of an EPN is decidable.  Our encoding represents the
configuration of an EPN by an ordered list of its marked places: thus
the configuration in Figure~\ref{fig:pcnet} is represented as the list
{\tt [b1,c1,p1]}.  Next, a transition $T$ is represented by a list of
places with input arcs to $T$ ($\bullet T$) and output arcs from $T$
($T \bullet$).  Predicate \texttt{trans/3} in Figure~\ref{fig:prog}
represents each transition of Figure~\ref{fig:pcnet} as a
fact, using XSB's trie indexing (see
Section~\ref{sec:dyn}) to obtain full indexing on list elements.
Figure~\ref{fig:prog} also shows code for determining reachability in
an EPN; for instance solutions to the goal {\tt
  reachable([b1,c1,p1],X)} are configurations reachable from the EPN
in Figure~\ref{fig:pcnet}.
%
\input{petri}
%For efficiency the reachability program assumes that the lists in all
%transitions and configurations are sorted.  
For a transition $T$ to have concession (be able to fire) in a
configuration $C$ of an EPN, every place in $\bullet T$ must be
marked, and no place in $T\bullet$ can be marked.  These conditions
are checked by 
%the predicate 
{\tt hasTransition/2} in Figure~\ref{fig:prog} which %recurses through
%the places in the current configuration ({\tt Conf}) to 
finds sets of
transitions that might have concession.  The recursion (in {\tt
  get\_trans\_for\_conf\_1/3}) allows indexed calls to transitions to
be made based on each place in the input configuration.  Each set of
possible transitions is then filtered to include only those
transitions that actually have concession
%in {\tt Conf},
 using operations on ordered sets (via {\tt
  check\_concession/2}).  {\tt hasTransition/2} succeeds when the
first of these transitions is applied; further transitions are applied
upon backtracking.

The predicate {\tt reachable/2} is a left-recursive reachability definition based on {\tt hasTransition/2}.
%Based on {\tt hasTransition/2}, a tabled reachability predicate is
%written as a simple left-recursion.  
Tabling {\tt reachable/2} is
useful in two ways: it prevents looping when a given configuration is
reachable from itself; and it filters out redundant paths to a
reachable configuration.  With the left recursive form of {\tt
  reachable/2} a typical call, such as {\tt reachable([b1,c1,p1],X)}
with first argument bound and second free, requires a single
tabled subgoal, and has as answers all configurations reachable
from {\tt [b1,c1,p1]}.  XSB's use of tries to represent tabled
subgoals and their answers allows efficient checking of answers and
efficient use of memory, since the trie data structure factors out
common list prefixes (cf. Section~\ref{sec:tries}).  Using this
program, nets with millions of transitions can be fully traversed in
under a minute.

%\end{example}

%\input{knapsack}

%\input{parsing}

\input{negation}

\input{system}

\input{callsub}

\input{anssub}

\subsection{Tabling with Constraints} \label{sec:constr}
XSB offers a simple integration of TLP and Constraint Logic
Programming (CLP) as follows.  XSB implements CLP by using attributed
variables, as do many other Prologs.  When an attributed (constrained)
variable $V_A$ is part of a tabled subgoal or a derived answer, $V_A$
is copied into the table along with its attributes; later when $V_A$
is copied out of a table, its attributes are also copied out and
associated with $V_A$.  For instance the query {\tt ?- p1(X)} to the
CLP(R) program
\begin{verbatim}
:- table p1/1.
p1(X):- {X < 9}.
\end{verbatim}
will return the answer {\tt \{ X < 9.0000\}}.  In \version{} of XSB,
attributed variables are supported even when they occur in literals
that are delayed, so that variables in a residual program may be
constrained.  Since entailment of constraints seen as a relation is a
partial order, answer subsumption can be supported for constrained
variables using the methods of the previous section.
%For constraint domains in which a join function is decidable
%answer subsumption using upper semi-lattices can also be
%used%~\footnote{In \version of XSB, constrained variables are not
%  allowed in subgoals of predicates tabled using call subsumption.}.

Many CLP applications will not benefit from tabling, particularly if
Prolog interacts with a constraint processor mainly to generate a set of
equations to be solved.  However for situations that require traversal
through a state space where states are associated with constraints,
tabling can be useful.  Tabled constraints have been used to analyze
security protocols~\cite{SarS05}, and for abstract interpretation
(cf.~\cite{CodF92} which pre-dates XSB) and when grammar rules
involve constraints (cf. \cite{Shie92}), tabled constraints can
provide the benefits to parsing. % described in  % fix ref.
%Example~\ref{ex:parsing}.
%Example 2.3.
The following example shows how tabled constraints can be used to
analyze a process logic.

{\em Example 2.6 (Constraint-Based Nets)}
A variety of formalisms extend Place Trans\-i\-tion Nets to add
conditions that must be evaluated for a transition to fire and to add
effects that must occur upon its firing, creating applications that
can be termed workflow nets.  A constraint net follows this model: 
%the Workflow nets of Example 2.5
%conditions and effects were Prolog predicates, but there is no reason
%why 
a condition is the entailment of a formula in a given constraint
domain, and the effect is the propagation of new constraints to
variables associated with given places in the net.  Using such an
approach, constraint-based reasoning can be incorporated into workflow
logics.
The top-level change to add constraints to the running example affects
the application of a transition to a configuration, in 
{\tt apply\_trans\_to\_conf/3} of Figure~\ref{fig:prog}:

\begin{tt} {\small
\begin{tabbing}
foo\=foooooooooooooooooo\=foooooooooooooooooo\=foo\=foooooooooooooooooooooooooo\=\kill
apply\_trans\_to\_conf(trans(In,Entailment,Out),Conf,NewConf):- \\
\>	unify\_for\_entailment(In,Conf,MidConf), \\
\>	entailed(Entailment), \\
\>	call\_new\_constraints(Out,OutPlaces), \\
\>	flatsort([OutPlaces|MidConf],NewConf).
\end{tabbing} }
\end{tt}
\noindent
First, variables in the transition are unified with those of the
configuration to produce a new constraint store.  If the formula {\tt
Entailment} is entailed by the constraint store, new constraints from
the transition are placed on the output variables via calling the
constraints in the list {\tt Out}.  Note that this extension is not
specific to a given constraint domain.
% but its use for reachability does depend on tabled constraints.

%% file: sld-tplp.tex
From a theoretical perspective SLD, the resolution method underlying
Prolog, is complete in that there is an SLD proof for every correct
answer for a query $Q$ to a program $P$.  However, the search
for an SLD proof may not terminate, even when $P$ is a datalog program.  For example,
consider {\tt ?- reach(1,Y)} to the program  $P_{Lrec}$:
% $P_{Lrec}$ in Figure~\ref{fig:sldtree}.  An SLD search tree for this
%query is also shown in Figure~\ref{fig:sldtree} where Prolog's
%top-to-bottom clause selection strategy and left-to-right literal
%selection strategy are both used.  Note that the tree contains proofs
%of (i.e. root-to-leaf paths to) both of the correct answer
%substitutions, $Y = 2$ and $Y = 3$, but that the tree is infinite and
%both of these answers lay to the right of an infinite branch.
\begin{verbatim}
reach(X,Y):- reach(X,Z),edge(Z,Y).           edge(1,2).  
reach(X,Y):- edge(X,Y).                      edge(2,3).
\end{verbatim}
An SLD search tree for this query provides proofs for both of the
correct answer substitutions, $Y=2$ and $Y=3$. 
However the SLD tree is infinite, and, when Prolog's search strategy is used,
both answers lie after an infinite branch.
I.e., Prolog will go into an infinite loop before
deriving the first answer.  Indeed, since the tree is infinite, no
complete search will ever terminate.
%
%\begin{figure}[htbp]
%\centering
%\epsfig{file=Figures/sldtree.eps,width=\textwidth}
%\caption{A program $P_{Lrec}$ and SLD tree for the query {\em reach(1,Y)}}
%\label{fig:sldtree}
%\end{figure}

%% file: petri.tex
% Program to determine reachability of an elementary net
\begin{figure}[htbp]
\begin{center}
{\small
\begin{verbatim}
1  :- table reachable/2.
   reachable(InConf,NewConf):-
      reachable(InConf,Conf),
      hasTransition(Conf,NewConf).
5  reachable(InConf,NewConf):- 
       hasTransition(InConf,NewConf).
      
   hasTransition(Conf,NewConf):-
      get_trans_for_conf(Conf,AllTrans),
10    member(Trans,AllTrans),
      apply_trans_to_conf(Trans,Conf,NewConf).
   
   get_trans_for_conf(Conf,Flattrans):-
      get_trans_for_conf_1(Conf,Conf,Trans),
15    flatten(Trans,Flattrans).
   
   get_trans_for_conf_1([],_Conf,[]).
   get_trans_for_conf_1([H|T],Conf,[Trans1|RT]):-
      findall(trans([H|In],Out,Tran),trans([H|In],Out,Tran),Trans),
20    check_concession(Trans,Conf,Trans1),
      get_trans_for_conf_1(T,Conf,RT).
   
   check_concession([],_,[]).
   check_concession([trans(In,Out,Name)|T],Input,[trans(In,Out,Name)|T1]):-
25    ord_subset(In,Input), 
      ord_disjoint(Out,Input),!,
      check_concession(T,Input,T1).
   check_concession([_Trans|T],Input,T1):-
      check_concession(T,Input,T1).
30
   apply_trans_to_conf(trans(In,Out_Name),Conf,NewConf):-
      ord_subtract(Conf,In,Diff),
      flatten([Out|Diff],Temp),
      sort(Temp,NewConf).

% Prolog representation of the Producer-Consumer Net
:- dynamic trans/2.
:- index(trans/2,trie).
trans([p1],[p2],t1).          trans([b2,p2],[p1,b1],t2).    
trans([b1,c1],[b2,c2],t3).    trans([c2],[c1],t4).    
\end{verbatim}
}
\end{center}
\caption{TLP Program for Analyzing Reachability of Elementary Petri Nets}
\label{fig:prog}
\end{figure}

%% file: negation.tex
\subsection{Tabled Negation} \label{sec:neg}
The following example illustrates 
%several aspects of 
evaluation of WFS using tabled negation in XSB.

%\begin{example} \label{ex:neg1}
{\em Example 2.3\ } %\label{ex:neg1}
Figure~\ref{fig:negprog} shows the normal program $P_{neg}$ where {\tt
  tnot/1} is XSB's predicate for tabled negation.  The atom {\tt p(c)}
is true in the well-founded model of $P_{neg}$ and the schematic
ground instantiation of $P_{neg}$ in Figure~\ref{fig:negproground}
illustrates why this is so.  First, {\tt p(c)} is true because {\tt
  p(a)} is false.  All except 2 of the 8 ground instances of clauses
for {\tt p(a)} are false because their first literal, a call to {\tt
  t/3} is false; the remaining two:
\begin{verbatim}
p(a) :- t(a,a,b), tnot p(a), tnot p(b).  
p(a) :- t(a,b,a), tnot p(b), tnot p(a).
\end{verbatim}
are false because {\tt p(b)} is true, so that {\tt tnot p(b)} is
false.

However in a tabled evaluation of {\tt p(a)} that uses Prolog's
literal selection strategy, the literal {\tt tnot p(a)} is selected
while evaluating the clause
\begin{verbatim}
p(a) :- t(a,b,a), tnot p(a), tnot p(b).
\end{verbatim}
leading to a loop through negation.  
%Of course if the evaluation could
%look ahead to see {\tt tnot p(b)}, it would immediately fail and the
%problem would go away.  
At this point, it might be tempting to try a
different search strategy, but 
it turns out that
%
%for $P_{neg}$ any search strategy that
%requires a fixed order of literal selection for all clauses will
%either flounder or will fail on the clause
%
%\begin{verbatim}
%p(a) :- t(a,a,b), tnot p(a), tnot p(b).  
%\end{verbatim}
%
%This problem is general -- 
no deterministic search strategy can
evaluate WFS top-down without encountering loops through negation.
The approach of SLG resolution is to {\em delay} the evaluation of a
literal involved in such a loop
% through negation 
and then to {\em simplify}
that literal later if it is determined to be true or false.

\begin{figure} 
\begin{verbatim}
:- table p/1.
p(b).
p(c) :- tnot p(a).
p(X) :- t(X,Y,Z), tnot p(Y), tnot p(Z).

t(a,a,b).       t(a,b,a).         
\end{verbatim}
\caption{A program, $P_{neg}$}
\label{fig:negprog}
\end{figure}

\begin{figure} 
\begin{verbatim}
p(b).
p(c):- tnot p(a).
p(a) :- t(a,a,a), tnot p(a), tnot p(a).  
p(a) :- t(a,a,b), tnot p(a), tnot p(b).  
:
p(a) :- t(a,b,a), tnot p(b), tnot p(a).
:
p(b) :- t(b,b,b), tnot p(b), tnot p(b).

t(a,a,b).        t(a,b,a).        
\end{verbatim}
\caption{The ground instantiation of $P_{neg}$}
\label{fig:negproground}
\end{figure}
% p(a) :- t(c,c,c), tnot p(c), tnot p(c).
%:
%p(c) :- t(c,c,c), tnot p(c), tnot p(c).
%t(c,c,c).

\begin{figure*}[htbp] 
\begin{center}
%\fbox{\epsfig{file=Figures/del-simpl3.eps,width=\textwidth}}
\epsfig{file=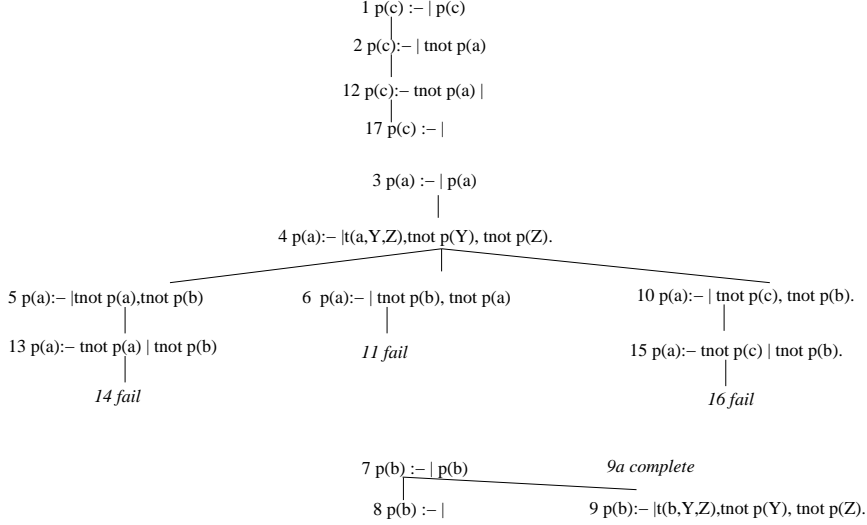,width=\textwidth}
\end{center}
\caption{SLG Evaluation of query {\tt ?- p(c)} to $P_{neg}$}
\label{fig:neg}
\end{figure*}

Figure~\ref{fig:neg} illustrates SLG resolution for this query and
program.  Within the nodes of Figure~\ref{fig:neg}, the new symbol
{\tt |} separates the unresolved goals to the right from the delayed
goals to the left.  In the evaluation state where nodes 1 through 10
have been created, {\tt p(b)} has been completed, and {\tt p(a)} and
{\tt p(c)} are in the same SCC.  There are no more clauses or answers
to resolve, but {\tt p(a)} is involved in a loop through negation in
node 5, and nodes 2 and 10 involve {\tt p(a)} and {\tt p(c)} in a
negative loop~\footnote{In this example, we ignore the effects of
  early completion which would complete {\tt p(b)} immediately upon
  creation of node 8, obviating the need to create node 9.}.

In situations such as this, where all resolution has been performed
for nodes in an SCC and where there are multiple literals that can be
delayed, an arbitrary one is chosen to be delayed first.  So the
evaluation delays the selected literal of node 2 to generate node 12
producing a {\em conditional answer} -- an answer with a non-empty
delay list.  Next {\tt tnot p(a)} in node 5 is delayed, failing that
computation path, and {\tt tnot p(c)} in node 10 is delayed to produce
node 15 and failing the final computation path for {\tt p(a)}.  At
this stage the computation of the SCC $\{p(a),p(c)\}$ is {\em
  completely evaluated} meaning that there are no more operations
applicable for goal literals.  Since {\tt p(a)} is completely
evaluated with no answers, conditional or otherwise, the evaluation
determines it to be false and a {\em simplification} operation can be
applied to the conditional answer of node 12, leading to the
unconditional answer in node 17.
%\end{example}

%Example~\ref{ex:neg1} illustrates several aspects of tabled negation
Example 2.3 illustrates several aspects of tabled negation in XSB.
First, the SLG operations of delay and simplification are used to
evaluate according to the WFS.  These operations are implemented at
the engine level, along with a mechanism for determining when tabled
literals may be involved in a negative loop.  Although delay and
simplification do not affect the complexity of SLG, it is inefficient
to perform them unnecessarily.
%Both Batched and Local evaluation are implemented in XSB so that they
%greatly reduce the number of delay operations needed; in fact 
XSB has
been implemented so that it is {\em delay minimal} for a large class
of programs~\cite{SaSW99}.  Second, the complexity of WFS is
reflected in the cost of operations in the example.  While a definite
program $P$ can be evaluated with abstract complexity $size(P)$, the
size of $P$, normal programs under WFS have abstract complexity
$atoms(P) \times size(P)$, the number of atoms in $P$ times the size of
$P$.  This complexity is reflected in an evaluation when XSB checks to
see whether loops of dependency are positive or negative: a check that
in the worst case might need to be done once per subgoal in an
SCC~\cite{SwPP09}.  Practically, these checks are usually performed
only once or twice per SCC, even when the SCCs are large and
non-stratified.  The result is that WFS evaluation usually scales
linearly with the size of a program: in fact it is difficult to
construct an example that scales with complexity $atoms(P) \times
size(P)$.

$P_{neg}$ has a 2-valued well-founded model, but in WFS the truth
value of atoms can be undefined.  From our tabling perspective, this
means that some conditional answers may have delayed literals that are
never simplified away, such as those in the program $P_{nonstrat}$:
\begin{verbatim}
:- table p/1, q/1.
p(1) :- tnot q(1).       q(1):- tnot p(1).
\end{verbatim}
%
%For this program, 
The query {\tt ?- p(X)} will succeed writing {\tt X = 1 undefined}
beneath the command-line prompt.  A call to {\tt
  get\_residual(p(X),D)} allows the conditional answer to be examined,
returning {\tt X = 1, D = [tnot(q(1))]}.  This highlights another
feature of SLG: it can be seen as a program transformation.  Given a
query $Q$ to a program $P$, XSB traverses that part of $P$ that is
relevant to proving $Q$, and creates its reduction with respect to the
well-founded model of $P$.  The tables produced by query evaluation
thus create a {\em residual program} that can be meta-interpreted
using {\tt get\_residual/2} or sent to a stable models solver through
XSB's XASP package.

XASP (XSB's Answer Set Programming package) provides several ways in
which information can be sent from XSB to an ASP solver.  \version{}
of XSB uses smodels~\cite{SiNS02}, which is included in XSB's
distribution, as its ASP solver.  The simplest way to use XASP is to
construct a partial stable model or 2-valued layered stable
model~\cite{PePi09} from the residual program.  For the preceding
program, the query {\tt ?- pstable\_model(q(1),Model,any)} will bind
{\tt Model} to each of the partial stable models for the residual
program of {\tt q(1)} -- in this case first to {\tt [p(1)]} and then
to {\tt [q(1)]}; {\tt ?- pstable\_model(q(1),Model,restrict)} will
restrict the models returned to those in which {\tt q(1)} is true.
Similarly the predicate {\tt in\_all\_stable\_models(Lit)} will
succeed if {\tt Lit} is true in all stable models, and there is at
least one stable model.  Using these routines, XSB can be used as a
query-oriented ASP grounder with both advantages and disadvantages
compared to grounders like {\tt lparse} and {\tt gringo}.  On the one
hand, cardinality or weight constraints, which are often used in
ASP~\cite{Niem99}, cannot be exploited if the residual program is sent
directly to smodels (although XASP has commands to send such
constraints separately to smodels).  Additionally, XSB may not be as
fast as a grounder like {\tt gringo} if a fully grounded program is
desired.  On the other hand, XASP is superior for grounding programs
that contain recursive data structures such as lists, for programs
where variables are instantiated over large domains, and for programs
and queries
%  in which a partial or layered stable model suffices so that 
where only partial grounding is required.

% end of comment

%% file: system.tex
\subsection{Implementation Aspects} \label{sec:impl}
So far tabling has been presented almost entirely through the forest
of trees model, which is sufficient for understanding many operational
aspects.  However, there are implementation aspects of XSB that are
useful in order to mix tabling with full Prolog, to understand and
control the space required by a tabled evaluation, and to write
programs that efficiently use XSB's tabling subsystem.

\vspace{-.1in}
\paragraph{Mixing Tabling and Prolog} 
SLD and SLG evaluation can be intermixed arbitrarily only if a
program does not contain side-effects or cuts.  A programmer should
take care when using a side effect for, say, I/O in a tabled predicate:
such a side-effect will be executed only the first time a given
subgoal is called, and not subsequently when the table is used.  The
behavior of cuts with tabled predicates requires 
explanation.  \version{} of XSB throws an exception if a computation
attempts to cut over a choice point for an incomplete table: that is,
a choice point that represents the root of an SLG tree or an internal
node with a selected tabled literal.  There is a semantic reason for
this.  Suppose a subgoal $S$ is called in two different places in the computation, place
1 with a cut and place 2 without.  If the cut for place 1 removed a
choice point of the above type, it could prohibit the derivation of
answers for place 2, and so give rise to incompleteness.  On the other
hand, XSB allows cuts over SLD choice points as well as cuts over
choice points for completed tables.
%In
%addition, XSB also allows cuts within the body of a tabled predicate:
%e.g., for a tabled predicate {\tt p/2} a clause such as
%\begin{verbatim}
%p(X,Y):- r(X,Y),!,...
%\end{verbatim}
%as long as the subgoal {\tt r(X,Y)} does not depend on an incomplete
%table.
\comment{\footnote{XSB allows this because its compilation of the cut
  creates a new non-tabled internal predicate, and choice points are
  cut to this predicate.}.
I removed this footnote, since I don't like the ``because'', and feel
it doesn't help understanding....(dsw)
}

\vspace{-.1in}
\paragraph{Implementing a Mechanism to Suspend and Resume} While
details of XSB's tabling engine, the SLG-WAM~\cite{SaSw98}, are beyond
the scope of this paper, we discuss a few aspects that are relevant to
its practical use.  First, the SLG-WAM implements the ability to
suspend and resume a computation by maintaining multiple computation
states within its environment stack, heap, choice point stack and
trail.  Whenever a binding is made to a trailed variable, that binding
is added to the trail frame itself, so that the SLG-WAM maintains a
{\em forward trail}.  Suspending and resuming are thus handled by
backtracking to unbind variables, and using the forward trail to
rebind the variables in a resumed path.  At various points in a tabled
evaluation, XSB freezes stack space so that the memory for suspended
computation paths is retained; stack space is reclaimed upon
completion of subgoals.  Thus, at a general level, the forest of trees
model maps to an implementation as follows: XSB associates each tabled
subgoal with its answers in a table.  Each non-completed tree, minus
its answers, is maintained in XSB's stacks, and its space is reclaimed
upon completion.  Heap space is also reclaimed by XSB's heap garbage
collector, which accounts for the multiple computation paths
maintained by the frozen stacks~\cite{DeSa01,CaSC01}.

Differences in the mechanism for suspending and resuming form the main
architectural differences in tabling engines.  YAP Prolog also implements the
SLG-WAM: its implementation is currently limited to definite programs,
but YAP Prolog also makes some important optimizations to the SLG-WAM to
improve speed~\cite{Roch01}.  Ciao Prolog implements a different
strategy, called CHAT~\cite{DeSa99a}, which suspends by copying (part
of) a computation path from the WAM stacks to a separate area of
memory, and resumes by copying the computation path back into the WAM
stacks.  CHAT can thus be thought of as an approach based on copying
rather than one based on sharing as with the SLG-WAM (performance
analysis in \cite{CaSW02} found the sharing approach to be superior
for many tabled programs).  B-Prolog uses a still different approach,
called linear tabling~\cite{ZhoS03}, which {\em rederives} a suspended
computation path rather than saving the suspended path in trail frames
or in a separate CHAT area.  
%Regardless of the actual engine design,
%it is best to complete subgoals as soon as they are determined to be
%completely evaluated.  At that point, only the subgoal and its
%(unique) answers need to be retained: the various computation paths
%leading to selected tabled subgoals can be released, leading to
%reclaimed stack space, reclaimed CHAT area, etc.

\vspace{-.1in}
\paragraph{Reclaiming Table Space} \label{sec:reclaim}

Tables factor subcomputations at the price of taking up space, so that
a practical system for tabled Prolog must provide a means to reclaim
the space that tables use.  XSB provides a number of predicates that
abolish table space safely, and that support different modes of
tabling.  Perhaps {\em query-level tabling} is the most common mode,
where tables ensure termination or a particular complexity for a user
query.  A second mode is {\em amortizing tabling} where tables reduce
the cost of multiple top-level queries by tabling repeated
subcomputations, even if these subcomputation are not repeated within
the same query.  A third mode is {\em user-controlled tabling} where
an application that uses tabling heavily decides itself when a table
is no longer needed and abolishes it, perhaps deeply within a
top-level query.  We offer a brief summary of the approaches taken to
support these modes.

Query-level tabling is perhaps the simplest use to support: if the
command-line interpreter (or a similar controller) calls the predicate
{\tt abolish\_all\_tables/0} at the end of a top-level query, all
tables are abolished and their space immediately reclaimed.
Furthermore, semantic problems with reclaiming space for incomplete
tables are avoided.  Nonetheless, XSB does not perform this by
default, since there are many reasons for maintaining tables between
user queries\footnote{XSB's command-line interpreter automatically
  reclaims space for any incomplete tables at the end of a query.
  Setting the Prolog flag {\tt query\_level\_tabling} ensures that all
  other tables are abolished as well.}.  For amortizing tabling, the
table space for certain predicates or groups of predicates often needs
to be reclaimed at once, possibly at the command-line, while tables
for other predicates should persist.  To support this, XSB provides the
predicate {\tt abolish\_table\_pred/1}, which abolishes tables for a
given predicate, and {\tt abolish\_table\_module/1}, which abolishes
tables for all predicates in a given module.  User-level tabling
sometimes requires a finer level of control, provided by the predicate
{\tt abolish\_table\_call/1} which abolishes a single table.
%, although
%the other table abolishing predicates can of course be used for
%user-level tabling.

In order to ensure the safety of user-controlled tabling, the system
must prevent the situation in which an evaluation abolishes a table,
reclaims its space, and then backtracks to a state that accesses the released structures.
To avoid this, XSB does not reclaim space for an
abolished table until there are no choice points pointing into that
table.  Instead, a {\em table garbage collector} periodically reclaims
space for abolished predicates and calls.  Any call made before the
abolish will be able to use answers from the abolished table.

The existence of residual programs further complicates space
reclamation.  In the program $P_{nonstrat}$ of Section~\ref{sec:neg},
suppose that the table for {\tt p(1)} were abolished but not that for
{\tt q(1)}.  If the residual program were traversed for
meta-interpretation or some other reason, the conditional answer {\tt
  q(1):- p(1)|} would point into the abolished table for {\tt p(1)},
To handle situations like this, when a table $T$ is abolished, XSB
ensures that other tables with conditional clauses that depend on $T$
or its answers are also (transitively) abolished.  Of course, some
applications may use the well-founded semantics but have no desire to
examine a residual program.  These applications can set the Prolog
flag {\tt table\_gc\_action} to override this behavior, so that
abolishing $T$ does not cause other tables to be abolished.

% multi-threaded engine?
% Finally, when multi-threading is used, space reclamation for each
% thread's private tables is exactly the same as that described above
% for a single thread, and the analogous predicates {\tt
%   abolish\_all\_private\_tables} and {\tt
%   abolish\_all\_shared\_tables} are provided.  In \version{} of XSB,
% however, space for shared tables may be reclaimed only when there is a
% single active thread.

\begin{figure} %[htbp]
%\centering
\begin{tabular}{ccc}
{\tt rt(a,f(a,b),a).} & 
{\tt rt(a,f(a,X),Y).} &  
{\tt rt(b,V,d).} 
\end{tabular}
%&
\begin{tabular}{c}
\epsfig{file=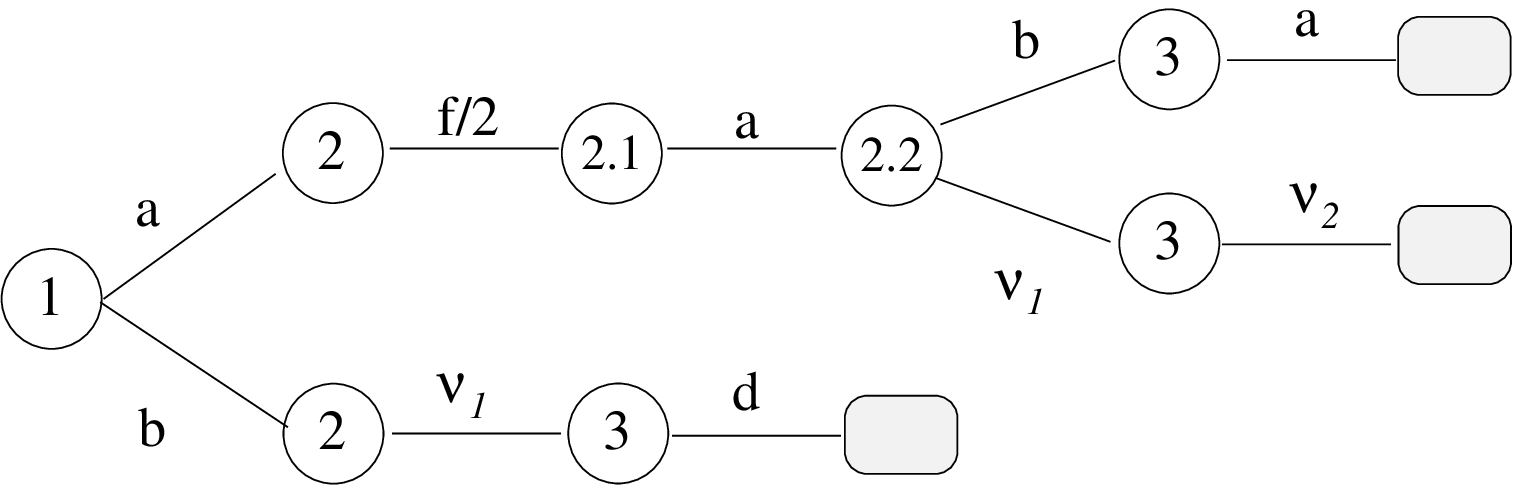,width=.7\textwidth}
\end{tabular}
%\\
%(a) & (b)
\caption{A set of terms and a schematic trie} \label{fig:trie}
\end{figure}

\vspace{-.1in}
\paragraph{Efficiently Accessing Tables} \label{sec:tries}
A data structure for tabling needs to
support three main operations: checking and/or inserting a new subgoal
in a table, checking and/or inserting a new answer in a table, and
backtracking through answers \cite{RRSSW98}.  A simple trie data
structure is well suited to support all of these operations.
Figure~\ref{fig:trie} depicts a set of terms along with a schematic
trie representing these terms.  The trie is built from a
prefix ordering of each term and thus in this case
factors out the common prefix of the first two terms.  Such factoring
has several advantages.  First, it can save space when sets of terms
have common prefixes.  Second, checking and inserting a term into a
trie can be done in one root to leaf pass, checking as long as the trie has a
symbol for the corresponding position of a term and inserting once a
position with no match is reached.  Third, backtracking through a
trie can be efficient, as common prefixes do not need to be untrailed
and rebound.  And fourth, in XSB each node of a trie is indexed, so
full term indexing is achieved.

In XSB, subgoals for a tabled predicate are kept in a subgoal trie,
and answers for each subgoal are kept in that subgoal's answer trie.
The overall structure is as if all tables for a given predicate had
been factored according to their subgoals.  These tables support an
optimization called {\em substitution factoring} which allows the
answer trie to contain only bindings.  For instance for a tabled
subgoal {\tt p(a,X,f(1,Y))} the answer trie would contain only the
bindings for {\tt X} and {\tt Y} and would not contain any constants
or structure symbols from the subgoal.  XSB also supports a {\em completed
  table optimization} in which the trie nodes themselves are SLG-WAM
instructions.  Thus backtracking through answers from a completed
table amounts to a direct execution of virtual machine instructions:
no meta-interpretation of the trie is needed. % to return such answers.

To summarize, a programmer can make use of XSB's table access by
writing programs where subgoals and answers can make use of the
left-to-right factoring provided by tries: such programs will also
backtrack through completed answers very quickly.  On the other hand,
substitution factoring ensures that when a subgoal has a large
structure, the structure needs to be traversed and stored only once
for the subgoal; subsequent answers pay no cost for the subgoal.

%% file: callsub.tex
\subsection{Call Subsumption} \label{sec:call-subs}
The preceding discussion of tabling was intentionally vague about
exactly how to determine whether a subgoal or answer is contained in a
given forest.  Given a selected atom $S$ and forest $\cF$, $S$ can
reuse the computation of a tree $S_{tab} \mif S_{tab}$ in $\cF$ as
long as $S_{tab}$ is at least as general as $S$ -- for our purposes,
as long as $S_{tab}$ subsumes $S$~\footnote{A term $T_1$ subsumes a term $T_2$ if there is
  an mgu of $T_1$ and $T_2$ whose domain consists only of variables of
  $T_1$.  If in addition the range of the mgu consists only of
  variables in $T_2$, $T_1$ and $T_2$ are variants.}.  Most implementations of tabling
only reuse tabled information if $S_{tab}$ is a {\em variant} of $S$,
but create a new tree otherwise: e.g. a subgoal {\tt p(a,X)} will
reuse the variant subgoal {\tt p(a,Y)}, but not the subsuming subgoal
{\tt p(X,Y)}.  In fact, the
evaluations in the preceding examples had no properly subsuming
subgoals, so their descriptions could ignore this distinction.

The distinction between these two approaches, call variance and call
subsumption, can radically affect the behavior of tabled evaluations.
Call subsumption can be especially useful when an entire model of a
target (sub-)program is desired -- as is the case in applications such
as program analysis, RDF inferencing or when combining rules with
ontologies.  This is because a fixed point can be initiated with a set
of open goals (goals without any bindings in their arguments); as
evaluation proceeds, tabled subgoals can reuse the answers from the
original set of subsuming subgoals.  For similar reasons, call
subsumption can be useful for deductive database-style queries that
are generated from a declarative framework as call subsumption is more
``forgiving'' when a program is not optimally written: for this reason
Flora-2 (see Section~\ref{sec:flora}) makes use of call subsumption
for certain generated queries.

When used in XSB, call subsumption allows a subgoal $S$ to use the
answers from a subgoal $S_{subsuming}$ as long as there is a table for
$S_{subsuming}$, regardless of whether $S_{subsuming}$ is completed or
not.  When $S_{subsuming}$ is computed along with several subsumed
non-completed subgoals, XSB's engine takes care to ensure that answers
are returned efficiently to the subsumed subgoals during the
iterations required to completely evaluate
$S_{subsuming}$~\cite{JRRR99}.  In \version{} call subsumption has
been extended to WFS, and supports answers with attributed variables
(to implement logical constraints), although subgoals must be free of
attributed variables~\cite{Swif09a}.  Either call subsumption or call
variance can be made the default for XSB, and a programmer can combine
them at a predicate level using the declaration:
%
%\begin{verbatim}
{\tt :- table p/n as <subsumptive/variant>.}
%\end{verbatim}
Call subsumption can provide substantial efficiency gains for many
programs; as a worst case, call subsumption in XSB imposes an overhead
of about 25\% for tabled evaluations that have no subsuming tabled
subgoals and so cannot make use of call subsumption.

{\em Example 2.4 (Querying Ontologies)\ }
%
%\begin{example}
An example of using call subsumption can be found in querying the 
OWL wine ontology ({\tt http://www.w3.org/TR/owl-guide}).  OWL
ontologies can be translated from RDF or HTML format into disjunctive
datalog by the KAON-2 system~\cite{Mot06}.  When translated by KAON-2,
the wine ontology produces a definite program with about 1000 clauses.
A small fragment of the translated code has the form:
\begin{verbatim}
pinotblanc(X) :- q24(X).
pinotblanc(X) :- pinotblanc(Y),kaon2equal(X, Y).
pinotblanc(X) :- wine(X),madefromgrape(X, Y),ot____nom21(Y).
madefromgrape(Y, X) :- madeintowine(X, Y).
madefromgrape(X, X) :- riesling(X),kaon2namedobjects(X).
madefromgrape(X, X) :- wine(X),kaon2namedobjects(X).
% 18 other clauses

wine(X) :- q14(X).                 q24(X) :- pinotblanc(X).
wine(X) :- texaswine(X).           q24(X) :- muscadet(X).
% 24 other clauses                 q24(X) :- q24(Y),kaon2equal(X,Y).
wine(X) :- q24(X).
% 31 other other clauses
\end{verbatim}
The generated program is highly recursive: for instance, {\tt
  pinotblanc(yellowTail)} depends on {\tt pinotblanc(X)} which depends
on {\tt wine(X)} and on {\tt wine(yellowTail)}.  In fact, nearly every
concept depends on nearly every other concept (more or less) due to
{\tt wine(X)} atoms occurring in body literals, and mixed
instantiations are often present in loops, due to the propagation of
bindings in rules.

Tabling can be used to evaluate a query to this ontology in XSB by
first using the {\tt :- auto\_table} declaration in the translated
file.  When XSB automatically tables, it chooses enough tables to break
all loops in the predicate dependency graph of a file or module.
Finding the minimal number of tables to break all loops is an
NP-complete problem, so that XSB uses a simple greedy algorithm.

When XSB evaluates the query {\tt pinotblanc(yellowTail)} using call
variance, it runs out of memory on a laptop machine.  When using call
subsumption to evaluate the query, XSB's time is comparable with the
ontoBroker system and is much faster than some ASP systems
\cite{LFWK09}.  Of course, WFS is not powerful enough to evaluate all
ontologies -- however for many translations XSB can create a residual
program that is passed to an ASP system using XASP.
%\end{example}

%% file: anssub.tex
\subsection{Answer Subsumption: Lattices, Partial Orders and Aggregation} \label{sec:ans-subs}
As with calls, there is a choice of whether to use answer variance or
answer subsumption: i.e., if a subgoal has answers {\tt p(a,Y)} and
{\tt p(a,b)}, does the evaluation use only the subsuming {\tt p(a,Y)}
for resolution with tabled subgoals, or does it use both answers?  XSB
and all other tabling Prologs use answer variance by default, as
answer subsumption between simple Prolog terms seems of limited use
for most programs.  However, if we generalize the notion of
subsumption from the lattice of terms to an arbitrary ordering $\cO$,
answer subsumption becomes quite useful~\cite{SwiW10a}.  Consider
first a case where $\cO$ is a partial order.  Such a case may prove
useful, for instance, in an inheritance hierarchy, where a query about
the relations of an object returns only the most specific answers.  In
this case, specific answers can be seen as the subsuming answers
according to the partial order of the inheritance hierarchy.

In addition to being a partial order, $\cO$ may also be a
lattice~\footnote{In \version{} of XSB only an upper semi-lattice need
  be defined.}: answer subsumption over lattices is extremely useful
for implementing paraconsistent and quantitative reasoning.

%\begin{example}
{\em Example 2.5 (Quantitative Degrees of Belief)}
Consider a model of quantitative degrees of belief~\cite{Emde86}.  An
annotated atom $A:[E_T,E_F]$ is an atom $A$ together with the annotation
\begin{itemize}
\item $E_T$, a number between 0 and 1 indicating 
%a measure of 
evidence that $A$ is true
\item $E_F$, a number between 0 and 1 indicating 
%a measure of 
evidence that $A$ is false.
\end{itemize}
In this model, $[E_T, E_F] \geq [E'_T,E'_F]$ if $E_T \geq E'_T$ and
$E_F \geq E'_F$, leading directly to a definition of a join operator:
\[ 
[E_T, E_F] \vee [E'_T, E'_F] = [ max(E_T,E'_T),max(E_F, E'_F)] 
\]
A ground atom $A:[E_T,E_F]$ is true in an interpretation $I$ of a
program $P$ if there are rules $A:[E^k_T,E^k_IF] :- Body^k$ in $P$
such that each $Body^k$ is true in $I$ and $\bigvee_{\forall k}
[E^k_T,E^k_F] \geq [E_T,E_F]$ in $I$.
%\end{example}

Writing programs over such a lattice is easy in XSB.  First, the
tabling declaration indicates that this lattice is to be used in a
particular argument of a tabled predicate.  The declaration:
\begin{verbatim}
:- table pred(_,_,qdb/3-[0,0]).
qdb([T1,F1],[T2,F2],[T3,F3]):-                 % join
    (T1 > T2 -> T3 = T1 ; T3 = T2),
    (F1 > F2 -> F3 = F1 ; F3 = F2).
\end{verbatim}
indicates that the third argument of {\tt pred/3} is to be maintained
via the lattice with join operation {\tt qdb/3} and identity (bottom)
{\tt [0,0]}.
{\tt pred/3} is translated into 
\begin{verbatim}
pred(X,Y,Z) :- bagReduce(Z1,pred_clause(X,Y,Z1),Z,qdb(_,_,_),[0,0]).
\end{verbatim}
where {\tt bagReduce/5} is a tabled predicate that performs the
answer subsumption. Each clause of {\tt pred/3} is translated into a
clause with head {\tt pred\_clause/3} where {\tt pred\_clause/3} is
non-tabled.  When the {\tt bagReduce} literal is
called, a check is made for a variant table.  If such a table is
not found, the subgoal creates a new table and calls clauses of {\tt
  pred\_clause/3} with a variable in the annotated position.  When a potential
answer, $pred(X_{ans},Y_{ans},Z_{ans})$ is derived, {\tt
  bagReduce/5} determines whether some answer exists whose first
two arguments are variants of $X_{ans}$ and $Y_{ans}$, with a third
argument $Z_{table}$.  If so, it takes the join $Z_{join}$ of
$Z_{ans}$ and $Z_{table}$.  If $Z_{join}$ is greater than $Z_{table}$,
the old answer is {\em deleted} from the table and the new answer
$pred(X_{ans},Y_{ans},Z_{join})$ is added.  The table for {\tt
  bagReduce/5} makes use of trie indexing for tables: the atom part
is earlier in the trie than the annotation part, which helps to make
the variant check of the atom efficient.  

\comment{
Partial orders are handled in an analogous manner.  Consider a
predicate {\tt desc(ItemID,TruncatedDesc)} that for each {\tt ItemID}
provides a set of possibly truncated descriptions of that item.  The
predicate {\tt extends/2} defines a partial order on strings, in which
a string is less than another if it is an extension of it:
\begin{verbatim}
:- table desc(_,extends/2).
extends(X,Y) :- append(X,_,Y).
\end{verbatim}
}

\version{} of XSB allows only positive recursion answer subsumption --
uses of negation must be stratified.  Non-stratified programs over
partial orders, can be modeled with preferences.  However, as shown
in~\cite{Swif99a} this method is sufficient to implement various
frameworks such as GAPs~\cite{KiSu92}, and Residuated
Programs~\cite{DamP01}.  In addition, it is easy to see that typical
aggregate functions, such as {\tt sum/3}, {\tt count/3}, {\tt min/3},
{\tt max/3} etc. are simple extensions of answer subsumption.

When answer subsumption is used, subsumed answers may be derived
before subsuming answers, so it is efficient to restrict those places
where non-maximal answers are used for resolution.  Local evaluation
is ideal for this, as it restricts all operations to a maximal SCC.
This property implies that no non-maximal answer will be used outside
of the SCC in which it was derived.  For this reason, the use of Local
evaluation can be critical for efficient answer subsumption:
\cite{FrSW98} provides an example where answer subsumption is used to
find the shortest paths in a graph $G$.  When Local evaluation is used
the time is proportional to the number of edges in $G$, but when
Batched evaluation is used, the time is proportional to the number of
{\em paths} in $G$, which is exponential in the number of edges of
$G$.

%% file: dynamic.tex
\section{Dynamic Code and Indexing} \label{sec:dyn}

%\subsection{Dynamic vs. Static Code}

\comment{
TLS: I think this is unnecessary -- after all, most of that is in the standard.

In XSB, as in many Prologs, a distinction is made between static and
dynamic code.  Clauses defining a predicate may be compiled from a
file, which generates a ``.xwam'' file containing object code, which
is then loaded as static code into the XSB emulator for execution.
Alternatively, clauses may be asserted at runtime (using one of the
variants of the {\tt assert/1} meta-predicate), in which case they
are directly compiled (with a simplified compilation strategy) to a
buffer in memory and linked into the memory structure to allow them to
be executed.  Dynamic code, i.e., clauses which are asserted, may
subsequently be removed (using one of the variants of the retract/1
meta-predicate.)
}

%Any Prolog that supports the core standard must implement dynamic
%code.  
In XSB, clauses for dynamic predicates are directly compiled with a
simplified compilation strategy, and dynamic predicates may be tabled.
While there are a variety of uses for dynamic code, perhaps the most
common use is for large and changing knowledge bases.  Dynamic code in
XSB supports a wider variety of indexing strategies than does static
code.  The performance of executing dynamic facts is comparable to
executing compiled facts with the same indexing, and with better
indexing strategies, dynamic code can be arbitrarily faster than the
corresponding static code.

%\subsection{Loading Large Fact Bases}
%
% TLS:  I dont think load_dyn/1 is a meta-predicate...
% TLS: load_dyn(c) handles multiple predicates in one file, I believe...
%
%Often applications require the loading into memory of large fact
%bases, i.e., a large number of fact clauses (clauses with a true body)
%for a single predicate.  
%
In XSB, large files of dynamic code can be efficiently loaded via {\tt
  load\_dyn/1} and its variants.  {\tt load\_dyn(File)} acts like a
compiler in recognizing directives, but treats all clauses in {\tt
  File} as dynamic; it re-consults {\tt File} by reading clauses from
{\tt File} and asserting them.
%~\footnote{{\tt load\_dyn/1} and its variants handle the {\tt
%    multi\_file} declaration.}.  
{\tt load\_dync/1} is a variation of {\tt load\_dyn/1} that can be
used if all clauses in a file are in (an extended) canonical form --
where no operators are used except for lists and comma-lists.  {\tt
  load\_dync/1} is extremely fast: as a general measure, it can read,
compile and load files of binary facts at a rate of about 300,000
facts per second on current hardware.
% TLS: just want to give order of magnitude

%\dsw{On my laptop, it takes about 3.4 seconds to load 1,000,000 binary
%  facts (of integers), and 5.7 seconds to load 1,000,000 binary facts
%  of unique atoms.} 

\comment{ Lots of systems have something like read_canonical, so I
  don't think we need much explanation here.  Also, I believe most systems
  implement assert so that it is about as fast as asserta.

XSB has a special reader that is invoked by the
builtin predicate {\tt read\_canonical/1}, which reads terms that are
in canonical form (i.e., do not use any operators), and is much faster
than the standard read.

uses {\tt read\_canonical} to read clauses from a file.
  Since in most cases, facts are in canonical form, {\tt load\_dync}
  is usually the best way to load large databases of facts in XSB.  A
  general {\tt load\_dyn/2} predicate supports several forms,
  including one in which {\tt asserta/1} is used to assert the facts,
  which can be useful when there are many hash collisions for an
  index.  \dsw{Terry suggests timings of load\_dync on various sized
    fact bases...}
}
%\subsection{Indexing of Dynamic Code} \label{sec:indexing}
\paragraph{Indexing of Dynamic Code} \label{sec:indexing}

Indexes for dynamic code are built using
\begin{itemize}
\item {\tt Trie Indexing} for which a trie is maintained to represent
  the entire predicate.  For instance, {\tt :-index(p/5,trie)} specifies
  trie-indexing for {\tt p/5}
\end{itemize}
or as combinations of
\begin{itemize}
\item {\tt Main-functor Indexing} for which a hash table is maintained
  for values of the main functor symbol of the indicated argument.
  For example {\tt :-index(p/5,3)} indexes the main functor symbol for
  the third argument of {\tt p/5}.
\item {\tt Star Indexing} for which a hash table is maintained for (up
  to) the first five symbols of the indicated argument.  Thus for
  example, star-indexing can distinguish the term {\tt [p(a)]} from
  the term {\tt [p(b)]}.  A declaration such as {\tt :-index(p/5,*(3))}
    star-indexes the third argument of {\tt p/5}.

\end{itemize}

Trie indexing is a special form of all-argument indexing, where
asserted facts are maintained in a trie (cf. Section~\ref{sec:tries}).
As with tabled answers and subgoals, the trie is built from a preorder
traversal of a fact, indexing at every position.  Clause ordering is
not maintained for trie indexed facts, and trie indexing cannot be
combined with any other indexing for a given predicate.  However
asserting and retracting to trie indexed code is about 3 times faster
than asserting or retracting to regular dynamic code.

Main functor and star indexing may be combined into multiple joint
indexes.  For example {\tt :-index(p/5,*(1)+3)}
%\begin{verbatim}
%:- index(p/5,*(1)+3).
%\end{verbatim}
asks for a joint index to be built (for future asserts to {\tt p/5})
for the first and third arguments, so that if a call is ground on both
its first and third arguments, it will index the indicated symbols
together. % in a single hash index. 
%\tls{single hash table?  What about
%the star?} \dsw{Yes, for one index, it's all just one hash-table.  All
%these declarations do is say what parts of the subgoal to use to
%construct the hash value.  Multiple different indexes need multiple
%different hash tables.}
Joint indexes may use up to three arguments.  Using this
joint index in a multiple index, the declaration {\tt :-index(p/5,[*(1)+2,*(1)])}
%\begin{verbatim}
%:- index(p/5,[*(1)+2,*(1)]).
%\end{verbatim}
causes two indexes to be built.  When calling {\tt p/5}, the indexes
are tried in left-to-right order.  For this example, if the first and
second arguments of a call are bound, the index {\tt *(1)+2} is used.
If argument 1 is bound but not argument 2, then {\tt *(1)} will be
used.

%\tls{A good example of the use of indexing can be seen in XSB's
%  ontology management package...}

\comment{
 not all those symbols are bound, then it will
see if just the first argument of the call can use the declared star
index, and if so, that will be used.  (Note that putting these two
indexes in the opposite order would result in the second,
multi-argument index never being used, since if the first fails to
apply, the second one will also.)
}
%\subsection{Incremental Table Maintenance}
\paragraph{Incremental Table Maintenance}
By default in XSB, tables are created when tabled goals are called and
are used until they are abolished.  But if a tabled predicate depends
on a dynamic predicate and the dynamic predicate changes, the table
becomes out of date.  This is known as the {\em view maintenance}
problem in databases and as the {\em truth maintenance} problem in
artificial intelligence.  
% Historically, XSB users were expected to be
% aware of dependencies of tables on changing dynamic predicates and
% explicitly abolish the out-of-date tables in such circumstances.
%However 
XSB provides support for {\em incremental tabling}, so that when
changes are made to dynamic predicates, dependent tables are
automatically updated to contain the corrected values~\footnote{
% In  \version{} of XSB, 
Currently incremental tabling is implemented only for call
  variance and for stratified programs.}.  Incremental tabling is
declared as: \verb|:- table p/2 as incremental|.
%\begin{verbatim}
%:- table p/2 as incremental.
%\end{verbatim}
To make use of incremental tabling, any dynamic predicate, such as
{\tt q/2}, whose change should trigger incremental table maintenance
is declared as: \verb|:- use_incremental_dynamic q/2|.
%\begin{verbatim}
%:- use_incremental_dynamic q/2.
%\end{verbatim}
In \version{} of XSB, incremental updates to a table can be triggered
in different ways.  In order to update a table based on a single
change to a dynamic predicate, calls such as {\tt
  incr\_assert(q(a,5))} or {\tt incr\_retract(q(a,5))} can be used.
For bulk changes to dynamic predicates, calls to {\tt assert/1} or
{\tt retract/1} are made, which will not trigger updates to tables.
At the end of the bulk ``transaction'' a call such as {\tt
  incr\_table\_update} triggers the appropriate updates.  Finally,
calls such as {\tt incr\_assert\_inval/1} {\em invalidate} tables that
depend on a dynamic predicate, for those cases where incremental
updates are deemed to be inefficient (e.g. a clause for a dynamic
tabled predicate is retracted).

% \tls{David: some readers may find this unexciting without an
%   appliction.  It would be great to include a paragraph about how you
%   use incremental tabling for XJ or for the deductive spreadsheet
%   here, or include it in the application section with a reference
%   here.}
% \dsw{I'll add something on deductive spreadsheets.  Incremental
% tabling is not used in XJ.  It should be, but the caching code for XJ
% was written before we had them, and I've never had time to go back and
% rewrite it.  I've been wanting to do that for some time, but XSB,
% Inc. can't support it, and I haven't found other time...}

A Deductive Spreadsheet system \cite{dss-ker} has been built by
programming an MS Excel plug-in in XSB to support recursive set
expressions in spreadsheet cells.  When a spreadsheet user updates the
contents of a cell, the engine must update the values of all cells
that depend on the updated cell.  This is an ideal application for
incremental table maintenance since the values of cells often depend
on the values of a few other cells, so most cells are not affected by
an update to some particular cell.  The implementation uses
incremental tabling to determine exactly the affected cells and then
updates only them.  Without incremental tabling, the plug-in was
limited to spreadsheets with a few hundred cells; with incremental
tabling the plug-in became practical and could recompute tables for
very large spreadsheets almost instantaneously.

%This application would not have been practical without incremental tabling: %For large tables most changes are computed almost
%%instantaneously.  Without incremental tabling, DSS would not be usable
%for spreadsheets with more than a few hundred cells.

%For large tables most changes are computed almost
%instantaneously.  Without incremental tabling, DSS would not be usable
%for spreadsheets with more than a few hundred cells.
%\tls{Can you be more specific about the savings?}
%\dsw{I don't have numbers.  Some examples were from many seconds down to an eyeblink.}

\comment{
In
this way every update to {\tt q/2} will trigger the update of all
tables that depend on it, and the tables will continue to have their
currently correct contents.

In some circumstances many updates are done to an {\tt
  incremental\_dynamic} predicate at once and there is no need to
maintain the correctness of dependent tables between each individual
change but only after all the changes have been made.  Lower-level
predicates are provided to support this situation.
}

%% file: mt.tex
\section{Multi-threading} \label{sec:mt}
Multi-threading, the ability to concurrently perform multiple computations, allows
Prolog to be used as a server that handles multiple
requests or as an agent that handles multiple types of input from its
environment.  The further ability to coordinate these computations
provides support for various types of parallel and distributed processing of a single
query.  
%Towards these ends, a number of Prologs have begun to support
%multi-threading.  
When multi-threading is combined with tabling and specialized
indexing, Prolog acquires functionality similar to that of a deductive
database, so that it can support applications from semantic web
reasoning systems to interactive GUI control.

XSB has been multi-threaded since version 3.0 and supports a draft
standard for multi-threaded Prologs (ISO/IEC DTR 13211-5:2007 2007).
%along with SWI-Prolog, YAP Prolog and others.  
The predicates in this standard, many of which originated in
SWI-Prolog~\cite{Wiel03}, provide facilities for various operations.
Predicates for creating, joining and exiting threads and for handling
mutexes provide a high-level interface to Prolog threads under a
Posix-style semantics.  Coordination among threads is handled by
message queues, which are used to pass Prolog terms among
threads. These message queues may be {\em public}, with multiple
readers and writers; or {\em thread-specific}, associated with a
thread that is the queue's only reader.  Among other uses,
thread-specific message queues form the basis for thread signaling,
which allows one thread to send a goal to another thread.  Threads
check for signals frequently, so that signaling becomes a powerful
mechanism for fine-grained, interrupt-based coordination.

\comment{ Signaling Prolog
  threads, however, is implemented within a Prolog engine, and so may
  have different behavior than Posix signaling  ~\footnote{When XSB is
    executing WAM code, signals are checked when attributed variables
    are checked (at beginning or end of a predicate's execution,
    before cuts and before adding a subgoal or answer to a table).
    However, \version{} does not guarantee that an XSB thread will be
    awakened when it is suspended for I/O or other reasons -- although
    XSB threads do handle signals when they have been put to sleep via
    {\tt thread\_sleep/1}.}.  \dsw{Is this level of detail necessary?}
}

\version{} of XSB may be configured either as single- or as
multi-threaded.  On Linux and Windows, Prolog evaluation is about
5-15\% slower for the multi-threaded engine than for the
single-threaded engine; however for Mac OS X the multi-threaded engine
is about 5-10\% faster.  The interface used to call C from XSB
supports both single- and multi-threading, so nearly all of XSB's
libraries and packages work under both engines.  In addition, both
engines can be embedded in C code.  When multi-threading is used, any
C thread can query any XSB thread that is not in the midst of a query.

All of these features form the basis of XSB's multi-threaded tabling
engine~\cite{Marques07}, which allows a thread to use private tables
to support an independent query, along with shared tables for subgoals
that may be used repeatedly by different threads.
The simplest execution model is that of private tables, where each
thread keeps its own copy of tabled information.  Private tables offer
several advantages:
\begin{itemize}
\item Private tables use sequential tabling algorithms
%  The main
%  implementation issues are to make the tabling engine reentrant with
%  a low overhead, to allow each thread to reclaim its own table space
%  and to ensure that allocation of table space does not affect
%  scalability.  Private tables in XSB support 
so that they naturally support all tabling features including tabled
negation, tabled constraints, and call and answer subsumption.
% TLS: need to get things working for incremental tabling.
%
\item Private tables generally require no synchronization among
  threads above the level of memory allocation.
\item Private tables are suitable to ensure query completeness or to
  support a particular semantics.  Tables are automatically reclaimed
  when the thread that computed them exits.  This reclamation includes
  not only subgoal and answer tries, but the delay lists and
  supporting structures used to compute WFS.
\end{itemize}

Shared tables are also important:
\begin{itemize}
\item  If different threads require the same tables, memory usage for
  shared tables will be significantly lower than for private tables.
\item Shared tables allow the decomposition of a program, so that a
  set of threads can together compute a set of tables, partially
  supporting Table-Parallelism \cite{FHSW95}.
\end{itemize}

{\bf Execution Models for Shared Tables}
%
%In \cite{RFM-PHD} two models for shared tables, {\em Concurrent Local
%  Evaluation} and {\em Concurrent Batched evaluation} were proposed
%and implemented. 
By default when tables are shared in \version, a model called {\em
  Concurrent Local evaluation} is used, which relies on Local
evaluation and dynamically partitions tables among threads.  The 
idea behind Concurrent Local evaluation is that when a thread $T$
encounters a (shared) tabled subgoal $S$ that has not been encountered
by any thread, $T$ evaluates $S$. Other threads are allowed to
use the table for $S$ only after $T$ has completed $S$.  Concurrency
control for tables mainly arises when more than one thread evaluates
different tabled subgoals in the same SCC at the same time.  In this
case, a deadlock will occur, which the engine detects and resolves, so
that a single thread assumes computation of all tabled subgoals in the
SCC~\cite{MarS08,MarS10}.  For example, in Figure~\ref{fig:local} such a
situation would occur if a thread $T_1$ called {\tt reach(1,Y)} and
another called {\tt reach(3,Y)} before it was called by $T_1$.  
%In \version{}, tabled subgoals that are computed by a new thread must
%have their answers recomputed.  It is shown in \cite{MarS08a} that
%recomputation does not add to the abstract complexity of WFS.

Because it is a type of Local evaluation, Concurrent Local evaluation
does not allow a consuming node to use answers produced by a subgoal
outside of its SCC until the table for the answers is completed -- a
restriction that prevents producer-consumer models of parallelism.
This limitation is overcome by {\em Concurrent Batched evaluation}
which allows several threads to compute (inter-)dependent tabled
subgoals in parallel.  As with Concurrent Local evaluation, each
subgoal can be computed by only one thread.  However, a given thread
may consume answers as they are produced by another thread.  The
implementation of Concurrent Batched evaluation extends the
implementation of sequential Batched evaluation.  In sequential
Batched evaluation, when the engine backtracks to the oldest subgoal
in an SCC, it schedules the return of unconsumed answers for each
consuming node in the SCC, and then proceeds to return the answers via
backtracking.
%by creating a chain of choice points, and
%then backtracks into the newly created chain.
%XSB's implementation of Concurrent Batched evaluation extends
%this strategy in a straightforward manner.  
%This approach is extended to a multi-threaded context as follows.  
In the multi-threaded context, if different threads compute different
SCCs, they can work independently, and can consume answers from other
threads as they become available.  However threads that run out of
work will suspend until a single thread institutes a fixpoint
check, after which the threads re-awaken.  Thus Concurrent Batched
evaluation allows parallel computation of subgoals, but has a
sequential fixpoint check that synchronizes multiple threads when they
compute the same SCC.

{\bf Implementation Status} 
%The status of multi-threading in XSB
%\version{} is shown in Table~\ref{tab:status}.  
In \version{} of XSB, private tables support all tabling features.
Concurrent Local evaluation supports most features, but does not yet
support call subsumption. 
% Also, as mentioned in
%Section~\ref{sec:reclaim}, it only partially supports space
%reclamation since shared tables can be abolished, but their space will
%not be reclaimed until there is only a single active thread in the
%engine.  
Both private tables and shared tables under Concurrent Local
evaluation have been heavily tested.  Concurrent Batched evaluation
%has been less tested and 
should be considered experimental and is currently restricted to
left-to-right dynamically stratified programs.

%Nonetheless, Concurrent Batched evaluation supports a number of
%tabling features, but 

%% file: applications.tex
\vspace{-.1in}
\section{Sample Applications of XSB} \label{sec:applic}
Numerous applications have been written using XSB: for space reasons
we restrict our discussion to two major applications.

{\bf XSB, Inc's Ontology-Driven Classification and Extraction}
Several large applications in XSB have been developed by the company
XSB, Inc.\footnote{XSB, Inc. (http://www.xsb.com) is a privately held
company that pursues applications of XSB and other technologies to
information retrieval and management.  XSB, Inc. has also helped
support the development of XSB and the related packages InterProlog
and XJ.}  Two important ones are the Ontology-Directed Classifier
(ODC) and the Ontology-Directed Extractor (ODE).
The ODC uses a modified Bayesian classification algorithm to classify
item descriptions to categories in a taxonomy.  It is in use quarterly
by the U.S. Department of Defense to classify over 80 million part
descriptions with respect to an extension of the UNSPSC taxonomy that
contains over 60,000 categories.  The ODE extracts attribute-value
pairs from those classified descriptions to build structured,
queriable knowledge about parts and their attributes.

Both depend critically on XSB.  First, the ontologies for both systems
are represented using XSB's CDF ontology management package.  CDF has
two facets: as a research system it supports the experimental
interaction of ontology axioms and rules as a hybrid MKNF knowledge
base (cf.\cite{GoAS10}).  However, for commercial use, it can simply
support the representation of classes with inheritance and typing,
objects belonging to those classes, and relationships among these
components.  It is very efficient for such ontologies, scaling to
gigabyte-sized in-memory ontologies and to even larger ontologies when
a relational database is used as a backing store.  CDF makes heavy use
of tabling (including tabled negation) along with the joint,
multiple, and star indexing discussed in Section~\ref{sec:indexing}.
Second, the applications perform a significant amount of text
processing using an XSB super-tokenizer module.  This module supports
the declaration of complex rewriting rules for token lists: tens of
thousands of these rules implement abbreviations and token corrections
in ODC and complex pattern-matching rules in ODE.  The super-tokenizer
uses tabled grammars and trie-based indexing in fundamental ways, as
well as negation to implement preferred rewritings.  
% as in preference grammars.
%
Third, both applications have complex training interfaces which allow
a knowledge expert to add extraction or classification rules,
experiment with how extraction or classification works on sample data,
and add information to improve the processes.  These interfaces are
built with XJ ({\tt www.xsb.com/xj.aspx}), an open-source package that
allows XSB to construct complex graphical user interfaces through the
Java Swing library.  XJ itself uses the open-source InterProlog
interface~\cite{Cale04} to communicate efficiently between Java and
XSB.

\begin{figure}
\centering
\epsfig{file=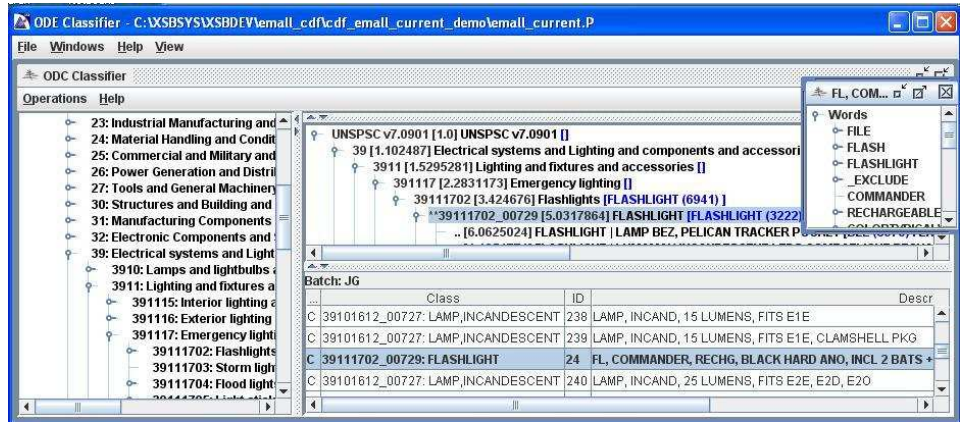,width=\textwidth}
\caption{A Screenshot of XSB Inc.'s Ontology-Directed Classifier}
\label{fig:odc}
\vspace{-.1in}
\end{figure}

Figure \ref{fig:odc} shows a screenshot of the ODC training GUI driven
by XSB using XJ.  The left panel displays the taxonomy that is the
target of the classification, here the standard UNSPSC taxonomy,
extended with further categories from the Federal Cataloging System.
The lower right panel shows item descriptions that are automatically
classified to the taxonomy categories.  The upper right panel shows
classification weights and thus an ``explanation'' of how the selected
description was classified as it was.  The optional floating window
in the upper right shows the words used in the classification after
abbreviation expansion and other rewriting.  A knowledge expert uses
this interface to explore how the classifier assigns descriptions and
to modify it by adding abbreviations, training items, and other tuning
options.

%\subsection{Flora-2 and Silk} \label{sec:flora}
{\bf Flora-2 and Silk} \label{sec:flora}
Flora-2~\cite{YaKZ03} ({\tt flora.sourceforge.org}) is a programming
system supporting Frame Logic (F-Logic)~\cite{KiLW95} HiLog and
Transaction logic, all of which are implemented in XSB.  Flora-2 is a
higher-level language than Prolog in the sense that it may represent
knowledge more concisely than Prolog, although it offers less
procedural control.

%\begin{example} \label{ex:flora}
{\em Example 5.1 (Flora-2)\ }
Figure~\ref{fig:flora} shows a fragment of a publications knowledge
base written in Flora-2.  This example was used in \cite{KiLW95} to
explain various features of F-logic; here we use it to briefly give a
flavor of Flora-2.  First, note that Flora-2 has a different syntax
from Prolog, although each of the statements is a well-defined term
and unification can be performed on these terms.  In addition, the
fragment is divided into a schema and its objects.  The subclass
relation is indicated by {\tt ::/2} and class membership for an object
by {\tt :/2}.  A class or object is associated with a set of its
attributes through brackets ({\tt []}).  Within a schema, class
attributes are indicated by {\tt =>>/2}, and by {\tt =>/2} if the
attributes are functional.  Inheritance for these attributes is
monotonic, that is each subclass inherits any concrete attributes of
its classes and super-classes and attributes may not be over-ridden.
Other predicates provide for inheritable attributes that may be
over-ridden.

\begin{figure} 
{\em Schema:}
{\small
\begin{verbatim}
conference_paper :: paper.
journal_paper :: paper.
paper[authors =>> person, title => string].
journal_paper[in_vol => volume].
conf_p[at_conf => conference_procs].
journal_vol[of => journal, volume => integer, number => integer, 
            year => integer].
journal[name => string, publisher => string, editors =>> person].
conference_procs[of_conf => conf_series,year => integer,editors =>> person].
conference_series[name => string].
publisher[name => string].
person[name => string, affiliation(integer) => institution].
institution[name => string, address => string].
\end{verbatim}
{\em Objects:}
\begin{verbatim}
o_j1 : journal p[title -> 'Records, Relations, Sets, Entities, and Things',
                 authors ->> {o_mes}, in vol -> o_i11].
o_di : conference_paper[title -> 'DIAM II and Levels of Abstraction',
                        authors ->> {o_mes, o_eba}, at_conf -> o_v76].
o_i11 : journal_vol[of -> o_is, number -> 1, volume -> 1, year -> 1975].
o_is : journal[name -> 'Information Systems', editors ->> {o_mj}].
o_v76 : conference_procs[of -> vldb, year -> 1976, 
                         editors ->> {o_pcl, o_ejn}].
o_vldb : conference_series[name -> 'Very Large Databases'].
o_mes : person[name -> 'Michael E. Senko'].
o_mj : person[name -> 'Matthias Jarke', affiliation(1976) -> o_rwt].
o_rwt : institution[name -> 'RWTH Aachen'].
\end{verbatim}
}
\caption{A Publications Object Base and its Schema in Flora-2} \label{fig:flora}
\vspace{-0.1in}
\end{figure}
%\end{example}

In addition to the features shown above
%Example~\ref{ex:flora}
inheritance and attribute predicates can be also defined in terms of
rules.  When deriving the answer to a query of a Flora-2 knowledge
base, resolution is performed as in Prolog, but the derivation also
makes use of inherited attributes and these attributes can be based on
other rules as can be the inheritance hierarchy itself. Thus a Flora-2
knowledge base has the advantages of an inheritance-based system for
knowledge representation.  The price it pays for this is the need to
traverse a potentially large portion of an inheritance hierarchy when
answering a query.  Tabling is a natural mechanism to factor out
subcomputations involving the inheritance hierarchies of objects, and
Flora-2 makes heavy use of tabling.  Flora-2 also relies on tabled
negation under WFS for non-monotonic inheritance.  The intuition
behind this is that an object non-monotonically inherits an attribute
if that attribute is {\em not} over-ridden by some other inherited
attribute.  Hierarchies with a well-defined ``over-rides'' relation
are stratified, but inheritance may be undefined in WFS.  For
instance, answers to the query {\tt ?- nixon[policy *-> X]} will be
undefined in the well-known ``Nixon Diamond'' example:
\begin{verbatim}
republican[policy *-> nonpacifist].     quaker[policy *-> pacifist].
nixon:republican.                       nixon:quaker.                          
\end{verbatim}

Flora-2 programs are compiled into XSB using a sophisticated series of
transformations.  These transformations decide what (XSB) predicates
need to be tabled, and also determine situations in which space can be
reclaimed making Flora-2 an example of {\em user-controlled tabling}
as discussed in Section~\ref{sec:reclaim}.  In many programs, a
hierarchy may be repeatedly traversed using calls in different modes,
so that the current experimental version of Flora-2 makes use of call
subsumption.  In addition, the Flora-2 compiler makes heavy use of
XSB's trie-indexed dynamic facts (Section~\ref{sec:indexing}) to
represent object code.
While it is a logic programming language, a Flora-2 program is
substantially different from a Prolog program.  Accordingly Flora-2
used Prolog to implement its own command-line interpreter, debugger
and module system rather than using those of XSB.  
%%Flora-2 also
%%implements HiLog (Section~\ref{sec:hilog}) as a transformation into
%%Flora-2 code that is then compiled into XSB.

The advantages of Flora-2 and XSB have given rise to its use in the
ambitious Digital Aristotle project ({\tt www.projecthalo.com})
described as ``a reasoning system capable of answering novel questions
and solving advanced problems in a broad range of scientific
disciplines and related human affairs.''  Digital Aristotle is based
on an extension of Flora-2 called Silk~\cite{Gros09} that contains
further features of defeasible reasoning and belief
logic~\cite{WGKFL09}, and which is implemented using the techniques of
the previous sections.

%\subsection{The ACORDA Agent Framework}~\label{sec:accorda}

%% file: discussion.tex
\section{Discussion}
%
%\tls{will need an update after the applications are written}

The various features discussed in this paper significantly expand the types of
programming that can be done in Prolog.  Tabling for definite programs
in itself allows sophisticated recursions to be coded simply and
efficiently; furthermore, these recursions can be combined with CLP as
shown in Section~\ref{sec:constr}.  The additions of tabled negation
and answer subsumption support a number of extensions such as
preferences and annotations;
and well-founded residual programs form a basis for
combining Prolog and ASP.  The use of call subsumption, incremental
tabling, and flexible indexing techniques for dynamic code supports
extensions of logic programs to deductive, object-oriented, and
semantic web databases -- this is particularly true when
multi-threading is also exploited.

Robust implementation of these 
%semantic and querying 
extensions have
led to a profusion of research and commercial applications, some of
which we cite here.  
%Interested readers are encouraged to consult the referenced papers.
Applications include those in program
verification~\cite{RRRSSW97,DuRS00,MRRV00,RRSDDRV00,KaTe02,PeRR02,PoRa04,SarS05},
%TLS: BKPR04,LiRS98, these look less application-oriented?
% TLS: DKRR98 -- two others like it already.
in program analysis~\cite{DRW96,Boul97,CoDS98,JaSa98,SaRa05}; in
natural language analysis and data
standardization~\cite{syntactica-semantica,RaRS97,RoLo98,CuSw02,DJPRSVW02},
in agent
implementations~\cite{ALPQ00,LeCK01,KaFi04,LGTH05,LTLH05,SaPe06} and
in semantic web
%ZFDCP0,
applications~\cite{PeAE98,DYHR00,LiPB02,TaDK03,SwiW03a,Swif04,ZoFC04,BhGr05,DrHM07},  
%AlKS09},
%ChFJ03
in diagnosis~\cite{CasP04,AABDS04,barata07}, in medical
informatics~\cite{GSTPD00,MuGR04}, in machine
learning~\cite{LARP00,PaGP01} and in software
%PeRo04
engineering~\cite{PeVi07,STICMC06,Oque04,dss-ker}. Many other
commercial applications have been developed by XSB, Inc., Medical
Decision Logics, Inc ({\tt www.mdlogix.com}), Ontology Works ({\tt
  www.ontologyworks.com}) and other companies.

%Different tabling applications have been explored by other Prolog
%systems.  The use of parallel tabling in YAP Prolog has been explored for
%verification~\cite{Roch02}, and YAP Prolog's tabling has been used for
%machine learning~\cite{ROFS05} and program analysis~\cite{BenF07}.
%Similarly, B-Prolog has formed the basis of probabilistic Prolog
%applications through PRISM~\cite{Sato09}.  
All of these applications demonstrate that TLP is a vibrant field of
research, involving numerous Prologs including XSB.

\vspace{-.1in}
\subsubsection*{Acknowledgements}
Dozens of people have contributed to the development XSB.  Among those
who have made contributions over a sustained period of time are (in
alphabetical order) Luis de Castro, Baoqiu Cui, Steve Dawson, Juliana
Freire, Ernie Johnson, Michael Kifer, Rui F. Marques, C.R. Ramakrishnsn, I.V. Ramakrishnan, Prasad Rao,
Konstantinos Sagonas and Diptikalyan Saha.  And we especially thank our
user community who have helped us find and fix so many bugs over the
years.